\documentclass[
]{ceurart}

\usepackage{soul}

\begin{document}

\copyrightyear{2022}
\copyrightclause{Copyright for this paper by its authors.
  Use permitted under Creative Commons License Attribution 4.0
  International (CC BY 4.0).}

\conference{18\textsuperscript{th} Italian Research Conference on Digital Libraries (IRCDL 2022)}

\title{Will open science change authorship for good? Towards a quantitative analysis}

\author[1]{Andrea Mannocci}[%
orcid=0000-0002-5193-7851,
email=andrea.mannocci@isti.cnr.it,
url=https://andremann.github.io,
]

\author[2,1]{Ornella Irrera}[%
orcid=0000-0003-2284-5699,
email=ornella.irrera@studenti.unipd.it,
]

\author[1,3]{Paolo Manghi}[%
orcid=0000-0001-7291-3210,
email=paolo.manghi@isti.cnr.it,
]


\address[1]{CNR-ISTI -- National Research Council, Institute of Information Science and Technologies ``Alessandro Faedo'', Pisa, Italy}
\address[2]{Department of Information Engineering, University of Padova, Italy}
\address[3]{OpenAIRE AMKE, Athens, Greece}

\begin{abstract}
Authorship of scientific articles has profoundly changed from early science until now.
If once upon a time a paper was authored by a handful of authors, scientific collaborations are much more prominent on average nowadays.
As authorship (and citation) is essentially the primary reward mechanism according to the traditional research evaluation frameworks, it turned to be a rather hot-button topic from which a significant portion of academic disputes stems.
However, the novel Open Science practices could be an opportunity to disrupt such dynamics and diversify the credit of the different scientific contributors involved in the diverse phases of the lifecycle of the same research effort.
In fact, a paper and research data (or software) contextually published could exhibit different authorship to give credit to the various contributors right where it feels most appropriate.
We argue that this can be computationally analysed by taking advantage of the wealth of information in model Open Science Graphs. 
Such a study can pave the way to understand better the dynamics and patterns of authorship in linked literature, research data and software, and how they evolved over the years.
\end{abstract}


\maketitle

\section{Introduction}
While, in early science, most of the papers were authored by a handful of scientists, modern science is characterised by more extensive collaborations, and the average number of authors per article has increased across many disciplines~\cite{cronin2001,wren2007write,baethge2008publish,fernandes2017evolution,frandsen2010}.
Indeed, in some fields of science (e.g., High Energy Physics), it is not infrequent to encounter hundreds or thousands of authors co-participating in the same piece of research~\cite{aad2012atlas}.
Such intricate collaboration patterns make it particularly hard to establish a correct relationship between contributor and scientific contribution, and hence get an accurate and fair reward during research evaluation~\cite{brand2015,vasilevsky2021}.
Thus, as widely known, scientific authorship tends to be a rather hot-button topic in academia as roughly one-fifth of academic disputes among authors stems from this~\cite{dance2012}.

Open Science, however, has the potential to disrupt such traditional mechanisms by injecting into the ``academic market'' new kinds of ``currency'' for credit attribution, merit and impact assessment~\cite{mooney2012,silvello2018}.
To this end, the new practices of research data (and software) deposition and citation could be perceived as an opportunity to diversify scientific attribution and eventually give credit -- right where it feels most appropriate -- to the different contributors involved in the diverse phases of the lifecycle within the same research endeavour~\cite{brand2015,bierer2017}.

In this extended abstract, we outline the perspective of using modern Open Science Graphs (OSGs) to analyse whether this is the case or not and understand if the opportunity has been seized already. 
Offering extensive metadata descriptions of both literature and research data records and the semantic relations among them, OSGs can be conducive to computational analysis of this phenomenon and thus study the emergence of significant patterns.
In particular, it will be interesting to analyse whether and how the authors' number, composition, and order varies when moving from literature to research data and software.

It would be, for example, interesting to discover that a significantly larger amount of people is involved in the development of software and the construction of datasets rather than in the editing of the related publications. 
This would confirm that the current reward mechanisms are obsolete and that there is a consistent, submerged workforce contributing to research that risks being underrepresented and under-evaluated if the current practices do not change for good.

Furthermore, modifications in the composition (by shuffle or by omission) of the authors participating in a publication as opposed to the ones contributing to related research data (or software) could instead reveal other interesting aspects worth investigating.
While, on the one hand, such changes in the two author lists could be contingent~\cite{kosmulski2012}, on the other, it could be interesting to relate them to the seniority of authors in order to detect patterns revealing a possible agency behind such a choice.
For example, data could suggest that senior staff members are less involved or, by any mean less interested, in participating, or getting rewarded, for data production and software development, thus confirming a bias towards the \textit{status quo} of research assessment.

\section{Data and methods}
\label{sec:methods}
In this section, we describe the dataset we intend to use to power our study, and we provide an overview of the methodology we intend to adhere to as well as the major caveats and challenges.

\subsection{Data}
The study here suggested is possible thanks to the ever-increasing amount of metadata about research products of the last decade.
In particular, Open Science Graphs (OSGs) can be a goldmine to this extent.
OSGs are Scientific Knowledge Graphs whose intent is to improve the overall FAIRness of science by enabling open access to graph representations of metadata about people, artefacts, institutions involved in the research lifecycle, as well as the relations between such entities, to support stakeholder needs, such as discovery, reuse, reproducibility, statistics, trends, monitoring, validation, and impact assessment. 
The represented information may span entities such as research artefacts (e.g., publications, data, software, samples, instruments) and items of their content (e.g., statistical hypothesis tests reported in publications), research organisations, researchers, services, projects, and funders. OSGs include relationships between such entities and sometimes formalised (semantic) concepts characterising them, such as machine-readable concept descriptions for advanced discoverability, interoperability, and reuse~\cite{aryani2020}.

For this analysis, we intend to adopt the OpenAIRE Research Graph\footnote{OpenAIRE Research Graph, \url{https://graph.openaire.eu}}~\cite{ManghiGraph} as our dataset of reference (hereafter, the Graph). 
The Graph is one of the core services provided by OpenAIRE AMKE\footnote{OpenAIRE, \url{https://www.openaire.eu}}, a not-for-profit legal entity operating an infrastructure that offers global services in support of Open Science scholarly workflows.
The Graph aggregates metadata from 96,514 scholarly sources (as of October 2021), comprising literature, research data and software repositories, publishers, and scholarly registries, such as ORCID, ROR, re3data, OpenDOAR, Crossref, and DataCite. It thus provides a longitudinal view of the global science record by delivering an extensive collection of heterogeneous research products interconnected with the relevant semantic relations. 
The semantic relations conducive to this study adhere to the specification drawn in the DataCite Schema documentation\footnote{DataCite schema, \url{https://schema.datacite.org}} and are both collected from DataCite\footnote{DataCite, \url{https://datacite.org}}, EMBL-EBI, and Crossref Event Data, as well as derived from the inference full-text algorithms embedded in the OpenAIRE Graph provision workflow, and the feedback from OpenAIRE portals users.

\subsection{Methods} 
First and foremost, a strategy to select the relevant literature and research data and software records needs to be devised.
To this end, let $p$ be a publication and $d$ a research data (or a software) contextually produced within the same research effort (e.g., $p$ describes a research effort to conduct a measuring campaign eventually producing the dataset $d$ released contextually to the publication).
In principle, it is possible to select all the $p \leftrightarrow d$ couples by looking at the semantics of the relations linking literature records to non-literature records within the OpenAIRE Research Graph.
In our case, we plan to use the semantic \textit{IsSupplementTo} (and inverse \textit{IsSupplementedBy}), which is the DataCite relation type indicating that a dataset $d$ is supplementary material for a publication $p$.

Once the relevant $p \leftrightarrow d$ couples have been selected, we need to proceed with the analysis of the author sets and their interpretation.
Let $A_p$ be the set of authors of a publication $p$, and $A_d$ the set of authors of a connected research data (or software) $d$, the scope of the analysis will be threefold.
Firstly, the cardinality of the two sets $|A_p|$ and $|A_d|$ can be compared to understand whether there is any difference in the workforce when moving from literature items to relevant research data.
Secondly, the composition of the two author sets will be considered in order to analyse the intersection $A_p \cap A_d$ and the symmetric difference $A_p\, \Delta \, A_d = (A_p \setminus A_d) \cup (A_d \setminus A_p)$.
Lastly, the ordering of the two author sets will be inspected to understand whether there are significant changes in the ranks of the same authors across the two sets and, if this is the case, which are the more frequent patterns.

Indeed, if the quantity of the available data points supports it, such analysis can be put in time perspective to analyse whether and how the trend evolved throughout the years globally and across different disciplines.

\subsection{Caveats}
This section analyses two identified major caveats and related challenges that we have to face in this analysis.
The first one is related to the inherent uncertainty of semantic relations specified among literature and research data records, while the second is related to the long-standing challenge of author disambiguation.
For each, we outline the strategy we intend to follow to solve, or at least mitigate, the side effects of the caveats here described.
    
\subsubsection{Semantic relation uncertainty}
Given a $p \leftrightarrow d$ couple, the semantic expressed by the relation is defined by the user (i.e., researcher, librarian, curator) taking care of the deposition process of the research product (e.g., on Zenodo).
Hence, the semantic is prone to human errors as it might not be very straightforward, which is the most appropriate one.
On Zenodo, for example, the choice is drawn from a dropdown menu with scarce or limited guidance on the rationale behind the choice.

In order to mitigate this aspect, we plan to run a heuristic over $p \leftrightarrow d$ couples tied by ``vanilla'' relations (e.g., \textit{Cites}, \textit{References}) and infer the unintentionally lost relations indicating supplemented material.
A viable strategy could consist in retrofitting as supplemented material relations all the \textit{Cites} (and inverse \textit{IsCitedBy}) and \textit{References} (and inverse \textit{IsReferencedBy}) relations when the author sets share at least an author and the year indicates that the two records are contextual (e.g., within six months apart).

A possible generalisation of the heuristic above would rely on multiple metadata fields such as the date of publication, the title and the author list itself to create a feature vector describing research outputs.
Then the distance between such vectors representing publication records and non-literature records related with the proper semantic would allow us to define a confidence interval of similarity which characterises literature and related non-literature records.
New supplement semantics can be inferred relying on such confidence interval: if the similarity between two feature vectors tied by ``vanilla'' relations lays within the interval, then the semantics has been probably misassigned, and thus it can be retrofitted as \textit{IsSupplementedBy} (or \textit{IsSupplementTo}, depending on the direction).

\subsubsection{Author names disambiguation}
Author disambiguation is essential to make the set of authors $A_d$ of the dataset (or software) $d$ and the set of authors $A_p$ of the publication $p$ comparable.

The metadata definition of an author $a$ who contributed both in the publication $p$ and in the supplement dataset $d$, may not be the same in $p$ and $d$ respectively. 
In this case, if the intersection $A_p \cap A_d$ is computed, the author $a$ will not belong to the intersection because there are different definitions of $a$ in $A_p$ and  $A_d$.
In this context, disambiguation is crucial to correctly detect that the author $a$ is the same in $A_p$ and $A_d$ despite multiple definitions. 

Consider, for example, the publication with the DOI:\url{https://doi.org/10.1186/s12865-015-0113-0} (\textit{Immune cell subsets and their gene expression profiles from human PBMC isolated by Vacutainer Cell Preparation Tube (CPT\textsuperscript{TM}) and standard density gradient}).
One of the datasets it is supplemented by is \url{https://doi.org/10.6084/m9.figshare.c.3600443_d4.v1} (\textit{Additional file 4: Table S4. of Immune cell subsets and their gene expression profiles from human PBMC isolated by Vacutainer Cell Preparation Tube (CPT™) and standard density gradient}). 
The lists of authors $A_p$ and $A_d$ are:
\begin{align*}
  A_p =\bigl\{{}& \text{Corkum, Christopher \textbf{P.}; Ings, Danielle \textbf{P.}; Burgess, Christopher;}\\ 
   &    \text{Karwowska, Sylwia; Kroll, Werner; Michalak, Tomasz \textbf{I.}}
    \bigr\}\\
  A_d =\bigl\{{}& \text{Corkum, Christopher; Ings, Danielle; Burgess, Christopher;}\\ 
   &    \text{Karwowska, Sylwia; Kroll, Werner; Michalak, Tomasz}
    \bigr\}
\end{align*}
If the lists of authors are analysed, there are three authors which co-occur both in $A_p$ and in $A_d$ (\textit{Burgess, Christopher}, \textit{Karwowska, Sylwia} and \textit{Kroll, Werner}).
The three remaining authors differ in how their \textit{name} is laid out: in $A_p$ in fact the first names are followed by another initial (highlighted in boldface), while in $A_d$ they do not; without finer author disambiguation strategies (e.g., plain string match) they would be considered different authors.

To address the author disambiguation problem, we can rely on the deduplication framework of OpenAIRE \cite{manghi2012duplication, manghi2020entity} and the distance metrics it provides to compute the distance between single authors and lists.
It is worth noting that, in contrast to the standard deduplication task (i.e., establish the equivalence of alike research products), we are comparing lists of authors belonging to research outputs different in kind (i.e., literature with non-literature); these lists may not necessarily contain the same authors; hence, the methods provided for the deduplication need to be customised according to our needs.


\begin{acknowledgments}
This work was co-funded by the European Commission H2020 project OpenAIRE-Nexus (grant number: 101017452). 
\end{acknowledgments}

\bibliography{biblio}

\begin{thebibliography}{17}
\expandafter\ifx\csname natexlab\endcsname\relax\def\natexlab#1{#1}\fi
\providecommand{\url}[1]{\texttt{#1}}
\providecommand{\href}[2]{#2}
\providecommand{\path}[1]{#1}
\providecommand{\DOIprefix}{doi:}
\providecommand{\ArXivprefix}{arXiv:}
\providecommand{\URLprefix}{URL: }
\providecommand{\Pubmedprefix}{pmid:}
\providecommand{\doi}[1]{\href{http://dx.doi.org/#1}{\path{#1}}}
\providecommand{\Pubmed}[1]{\href{pmid:#1}{\path{#1}}}
\providecommand{\bibinfo}[2]{#2}
\ifx\xfnm\relax \def\xfnm[#1]{\unskip,\space#1}\fi
\bibitem[{Cronin(2001)}]{cronin2001}
\bibinfo{author}{B.~Cronin},
\newblock \bibinfo{title}{Hyperauthorship: {{A}} postmodern perversion or
  evidence of a structural shift in scholarly communication practices?},
\newblock \bibinfo{journal}{Journal of the American Society for Information
  Science and Technology} \bibinfo{volume}{52} (\bibinfo{year}{2001})
  \bibinfo{pages}{558--569}. \DOIprefix\doi{10.1002/asi.1097}.
\bibitem[{Wren et~al.(2007)Wren, Kozak, Johnson, Deakyne, Schilling, and
  Dellavalle}]{wren2007write}
\bibinfo{author}{J.~D. Wren}, \bibinfo{author}{K.~Z. Kozak},
  \bibinfo{author}{K.~R. Johnson}, \bibinfo{author}{S.~J. Deakyne},
  \bibinfo{author}{L.~M. Schilling}, \bibinfo{author}{R.~P. Dellavalle},
\newblock \bibinfo{title}{The write position: A survey of perceived
  contributions to papers based on byline position and number of authors},
\newblock \bibinfo{journal}{EMBO reports} \bibinfo{volume}{8}
  (\bibinfo{year}{2007}) \bibinfo{pages}{988--991}.
\bibitem[{Baethge(2008)}]{baethge2008publish}
\bibinfo{author}{C.~Baethge},
\newblock \bibinfo{title}{Publish together or perish: the increasing number of
  authors per article in academic journals is the consequence of a changing
  scientific culture. some researchers define authorship quite loosely},
\newblock \bibinfo{journal}{Deutsches Arzteblatt International}
  \bibinfo{volume}{105} (\bibinfo{year}{2008}) \bibinfo{pages}{380}.
\bibitem[{Fernandes and Monteiro(2017)}]{fernandes2017evolution}
\bibinfo{author}{J.~M. Fernandes}, \bibinfo{author}{M.~P. Monteiro},
\newblock \bibinfo{title}{Evolution in the number of authors of computer
  science publications},
\newblock \bibinfo{journal}{Scientometrics} \bibinfo{volume}{110}
  (\bibinfo{year}{2017}) \bibinfo{pages}{529--539}.
\bibitem[{Frandsen and Nicolaisen(2010)}]{frandsen2010}
\bibinfo{author}{T.~F. Frandsen}, \bibinfo{author}{J.~Nicolaisen},
\newblock \bibinfo{title}{What is in a name? {{Credit}} assignment practices in
  different disciplines},
\newblock \bibinfo{journal}{Journal of Informetrics} \bibinfo{volume}{4}
  (\bibinfo{year}{2010}) \bibinfo{pages}{608--617}.
  \DOIprefix\doi{10.1016/j.joi.2010.06.010}.
\bibitem[{Aad et~al.(2012)}]{aad2012atlas}
\bibinfo{author}{G.~Aad}, et~al.,
\newblock \bibinfo{title}{Atlas collaboration},
\newblock \bibinfo{journal}{PHYSICAL REVIEW D Phys Rev D} \bibinfo{volume}{85}
  (\bibinfo{year}{2012}) \bibinfo{pages}{012003}.
\bibitem[{Brand et~al.(2015)Brand, Allen, Altman, Hlava, and Scott}]{brand2015}
\bibinfo{author}{A.~Brand}, \bibinfo{author}{L.~Allen},
  \bibinfo{author}{M.~Altman}, \bibinfo{author}{M.~Hlava},
  \bibinfo{author}{J.~Scott},
\newblock \bibinfo{title}{Beyond authorship: Attribution, contribution,
  collaboration, and credit},
\newblock \bibinfo{journal}{Learned Publishing} \bibinfo{volume}{28}
  (\bibinfo{year}{2015}). \DOIprefix\doi{10.1087/20150211}.
\bibitem[{Vasilevsky et~al.(2021)Vasilevsky, Hosseini, Teplitzky, Ilik,
  Mohammadi, Schneider, Kern, Colomb, Edmunds, Gutzman, Himmelstein, White,
  Smith, O'Keefe, Haendel, and Holmes}]{vasilevsky2021}
\bibinfo{author}{N.~A. Vasilevsky}, \bibinfo{author}{M.~Hosseini},
  \bibinfo{author}{S.~Teplitzky}, \bibinfo{author}{V.~Ilik},
  \bibinfo{author}{E.~Mohammadi}, \bibinfo{author}{J.~Schneider},
  \bibinfo{author}{B.~Kern}, \bibinfo{author}{J.~Colomb},
  \bibinfo{author}{S.~C. Edmunds}, \bibinfo{author}{K.~Gutzman},
  \bibinfo{author}{D.~S. Himmelstein}, \bibinfo{author}{M.~White},
  \bibinfo{author}{B.~Smith}, \bibinfo{author}{L.~O'Keefe},
  \bibinfo{author}{M.~Haendel}, \bibinfo{author}{K.~L. Holmes},
\newblock \bibinfo{title}{Is authorship sufficient for today's collaborative
  research? a call for contributor roles},
\newblock \bibinfo{journal}{Accountability in Research} \bibinfo{volume}{28}
  (\bibinfo{year}{2021}) \bibinfo{pages}{23--43}.
  \DOIprefix\doi{10.1080/08989621.2020.1779591}.
\bibitem[{Dance(2012)}]{dance2012}
\bibinfo{author}{A.~Dance},
\newblock \bibinfo{title}{Authorship: Who's on first?},
\newblock \bibinfo{journal}{Nature} \bibinfo{volume}{489}
  (\bibinfo{year}{2012}) \bibinfo{pages}{591--593}.
  \DOIprefix\doi{10.1038/nj7417-591a}.
\bibitem[{Mooney and Newton(2012)}]{mooney2012}
\bibinfo{author}{H.~Mooney}, \bibinfo{author}{M.~Newton},
\newblock \bibinfo{title}{The {{Anatomy}} of a {{Data Citation}}: Discovery,
  {{Reuse}}, and {{Credit}}},
\newblock \bibinfo{journal}{Journal of Librarianship and Scholarly
  Communication} \bibinfo{volume}{1} (\bibinfo{year}{2012})
  \bibinfo{pages}{eP1035}. \DOIprefix\doi{10.7710/2162-3309.1035}.
\bibitem[{Silvello(2018)}]{silvello2018}
\bibinfo{author}{G.~Silvello},
\newblock \bibinfo{title}{Theory and {{Practice}} of {{Data Citation}}},
\newblock \bibinfo{journal}{Journal of the Association for Information Science
  and Technology} \bibinfo{volume}{69} (\bibinfo{year}{2018})
  \bibinfo{pages}{6--20}. \DOIprefix\doi{10.1002/asi.23917}.
  \href{http://arxiv.org/abs/1706.07976}{{\tt arXiv:1706.07976}}.
\bibitem[{Bierer et~al.(2017)Bierer, Crosas, and Pierce}]{bierer2017}
\bibinfo{author}{B.~E. Bierer}, \bibinfo{author}{M.~Crosas},
  \bibinfo{author}{H.~H. Pierce},
\newblock \bibinfo{title}{Data {{Authorship}} as an {{Incentive}} to {{Data
  Sharing}}},
\newblock \bibinfo{journal}{New England Journal of Medicine}
  \bibinfo{volume}{376} (\bibinfo{year}{2017}) \bibinfo{pages}{1684--1687}.
  \DOIprefix\doi{10.1056/NEJMsb1616595}.
\bibitem[{Kosmulski(2012)}]{kosmulski2012}
\bibinfo{author}{M.~Kosmulski},
\newblock \bibinfo{title}{The order in the lists of authors in multi-author
  papers revisited},
\newblock \bibinfo{journal}{Journal of Informetrics} \bibinfo{volume}{6}
  (\bibinfo{year}{2012}) \bibinfo{pages}{639--644}.
  \DOIprefix\doi{10.1016/j.joi.2012.06.006}.
\bibitem[{Aryani et~al.(2020)Aryani, Fenner, Manghi, Mannocci, and
  Stocker}]{aryani2020}
\bibinfo{author}{A.~Aryani}, \bibinfo{author}{M.~Fenner},
  \bibinfo{author}{P.~Manghi}, \bibinfo{author}{A.~Mannocci},
  \bibinfo{author}{M.~Stocker},
\newblock \bibinfo{title}{Open {{Science Graphs Must Interoperate}}!},
\newblock in: \bibinfo{booktitle}{{{ADBIS}}, {{TPDL}} and {{EDA}} 2020 {{Common
  Workshops}} and {{Doctoral Consortium}}}, \bibinfo{publisher}{{Springer}},
  \bibinfo{year}{2020}, pp. \bibinfo{pages}{195--206}.
\bibitem[{Manghi et~al.(2020)Manghi, Atzori, Bardi, Baglioni, Schirrwagen,
  Dimitropoulos, La~Bruzzo, Foufoulas, Löhden, Bäcker, Mannocci, Horst,
  Jacewicz, Czerniak, Kiatropoulou, Kokogiannaki, De~Bonis, Artini, Ottonello,
  Lempesis, Ioannidis, Manola, and Principe}]{ManghiGraph}
\bibinfo{author}{P.~Manghi}, \bibinfo{author}{C.~Atzori},
  \bibinfo{author}{A.~Bardi}, \bibinfo{author}{M.~Baglioni},
  \bibinfo{author}{J.~Schirrwagen}, \bibinfo{author}{H.~Dimitropoulos},
  \bibinfo{author}{S.~La~Bruzzo}, \bibinfo{author}{I.~Foufoulas},
  \bibinfo{author}{A.~Löhden}, \bibinfo{author}{A.~Bäcker},
  \bibinfo{author}{A.~Mannocci}, \bibinfo{author}{M.~Horst},
  \bibinfo{author}{P.~Jacewicz}, \bibinfo{author}{A.~Czerniak},
  \bibinfo{author}{K.~Kiatropoulou}, \bibinfo{author}{A.~Kokogiannaki},
  \bibinfo{author}{M.~De~Bonis}, \bibinfo{author}{M.~Artini},
  \bibinfo{author}{E.~Ottonello}, \bibinfo{author}{A.~Lempesis},
  \bibinfo{author}{A.~Ioannidis}, \bibinfo{author}{N.~Manola},
  \bibinfo{author}{P.~Principe}, \bibinfo{title}{Openaire research graph dump},
  \bibinfo{year}{2020}. \URLprefix
  \url{https://doi.org/10.5281/zenodo.4279381}.
  \DOIprefix\doi{10.5281/zenodo.4279381}.
\bibitem[{Manghi et~al.(2012)Manghi, Mikulicic, and
  Atzori}]{manghi2012duplication}
\bibinfo{author}{P.~Manghi}, \bibinfo{author}{M.~Mikulicic},
  \bibinfo{author}{C.~Atzori},
\newblock \bibinfo{title}{De-duplication of aggregation authority files},
\newblock \bibinfo{journal}{International Journal of Metadata, Semantics and
  Ontologies} \bibinfo{volume}{7} (\bibinfo{year}{2012})
  \bibinfo{pages}{114--130}.
\bibitem[{Manghi et~al.(2020)Manghi, Atzori, De~Bonis, and
  Bardi}]{manghi2020entity}
\bibinfo{author}{P.~Manghi}, \bibinfo{author}{C.~Atzori},
  \bibinfo{author}{M.~De~Bonis}, \bibinfo{author}{A.~Bardi},
\newblock \bibinfo{title}{Entity deduplication in big data graphs for scholarly
  communication},
\newblock \bibinfo{journal}{Data Technologies and Applications}
  (\bibinfo{year}{2020}).

\end{thebibliography}

\end{document}